\documentclass[11pt]{article}

\usepackage{amsthm,amssymb,fullpage,times}

\newcommand{\ket}[1]{| #1 \rangle}
\newcommand{\bra}[1]{\langle #1 |}
\newcommand{\ketbra}[1]{\ket{#1}\bra{#1}}
\newcommand{\braket}[2]{\langle #1 | #2 \rangle}
\newcommand{\qform}[3]{\langle #1 | #2 | #3 \rangle}

\newcommand{\onemat}
           {\leavevmode\hbox{\small1\kern-3.8pt\normalsize1}}

\newcommand{\R}{\mathbb{R}}
\newcommand{\C}{\mathbb{C}}
\newcommand{\Z}{\mathbb{Z}}

\newcommand{\cA}{\mathcal{A}}
\newcommand{\cF}{\mathcal{F}}
\newcommand{\cM}{\mathcal{M}}
\newcommand{\cN}{\mathcal{N}}
\newcommand{\cP}{\mathcal{P}}
\newcommand{\cE}{\mathcal{E}}
\newcommand{\cG}{\mathcal{G}}
\newcommand{\cB}{\mathcal{B}}

\newcommand{\hcB}{\widehat{\mathcal{B}}}
\newcommand{\hsigma}{\hat{\sigma}}
\newcommand{\hb}{\hat{b}}
\newcommand{\hv}{\hat{v}}
\newcommand{\hG}{\widehat{G}}

\newcommand{\bsigma}{\bar{\sigma}}
\newcommand{\bcM}{\overline{\mathcal{M}}}

\newcommand{\tcB}{\widetilde{\mathcal{B}}}
\newcommand{\tb}{\tilde{b}}

\newcommand{\QFT}{{\rm QFT}}
\newcommand{\E}{\mathrm{E}}
\newcommand{\Tr}[0]{{\rm Tr}}
\newcommand{\rank}[1]{{\rm rank}(#1)}

\newcommand{\norm}[1]{\|{ #1 }\|}
\newcommand{\totvar}[1]{\|{ #1 }\|_1}
\newcommand{\trnorm}[1]{\|{ #1 }\|_{\mathrm{tr}}}
\newcommand{\frob}[1]{\|{ #1 }\|_{\mathrm{F}}}

\newcommand{\ceil}[1]{\left\lceil #1 \right\rceil}

\newtheorem{theorem}{Theorem}
\newtheorem{lemma}{Lemma}
\newtheorem{proposition}{Proposition}
\newtheorem{corollary}{Corollary}
\newtheorem{fact}{Fact}
\newtheorem{definition}{Definition}
\newtheorem{result}{Result}

\begin{document}
\title{Random measurement bases, quantum state distinction and
       applications to the hidden subgroup problem}

\author{
Pranab Sen \\
NEC Laboratories America, Inc. \\
4 Independence Way, Suite 200, \\
Princeton, NJ 08540, U.S.A.\\
{\tt pranab@nec-labs.com}
}

\date{}
\maketitle

\begin{abstract}
We show that measuring any two quantum states by a random POVM,
under a suitable definition of randomness, gives probability
distributions having total variation distance at least a universal
constant times the
Frobenius distance between the two states, with high probability.
In fact, if the Frobenius distance between the two states
is not too small and their ranks
are not too large, even a random orthonormal basis 
works as above.
Since a random POVM
is independent of the two states, the above result gives us
the first sufficient condition and an information-theoretic solution
for the following quantum {\em state distinction problem}:
given an a priori known ensemble of quantum states, is 
there a single
measurement basis, or more generally a POVM, that 
gives reasonably large total variation distance
between every pair of states from the ensemble?
Large pairwise trace distance is a trivial
necessary condition for the existence of a single distinguishing
measurement for an ensemble;
however, it is not sufficient,
as seen for example by the recent 
work of Moore, Russell and Schulman~\cite{MRS:sn}
on hidden subgroups of the symmetric group. 
Our random POVM method gives us the first 
information-theoretic upper bound on the number of copies required
to solve the quantum 
{\em state identification problem} for general ensembles, i.\,e., 
given some number of 
independent copies of a quantum state from an a priori known 
ensemble, identify the state. Moreover, this upper bound is
achieved by a {\em single register} algorithm, i.\,e.,
the algorithm
measures one copy of the state at a time, followed
by a classical post-processing on the observed outcomes in order
to identify the state. 

The standard quantum approach to solving the hidden 
subgroup problem (HSP), which includes Shor's algorithms for
factoring and discrete logarithm,
is a special case of the state identification problem where the
ensemble consists of so-called {\em coset states} of candidate
hidden subgroups. 
Combining Fourier sampling with our random POVM result
gives us single register
algorithms using polynomially many copies of the coset state that
identify hidden subgroups having
polynomially bounded rank in every representation of the ambient
group. In particular, we get such single register algorithms
when the hidden subgroup
forms a Gel'fand pair, e.g. dihedral, affine and Heisenberg groups,
with the ambient group, i.\,e.,
the rank in every representation is either
zero or one. 
These HSP algorithms complement earlier
results about the powerlessness of random Fourier sampling
when the ranks are exponentially large, which happens for example
in the HSP over the symmetric group.
The drawback of random Fourier sampling based 
algorithms is that they are not efficient because 
measuring in
a random basis is not. This leads us to the 
open question of
efficiently implementable pseudo-random measurement bases.
\end{abstract}

\section{Introduction}
The hidden subgroup problem (HSP) is a central problem in
quantum algorithms.
Many important problems like factoring, discrete
logarithm and graph isomorphism reduce to special cases of the HSP. 
Almost all exponential speedups that have been achieved in quantum
computing are obtained by solving some instances of the HSP. 
The HSP is defined as follows: Given a function 
$f : G \to S$ from a group $G$ to a set
$S$ that is constant on left cosets of some subgroup $H \leq G$ and
distinct on different cosets, find a set of generators for $H$.
Ideally, we would like to find $H$ in time polynomial in the input
size, i.\,e. $\log |G|$.
Almost all efficient quantum algorithms for solving 
special cases of the HSP,
including Shor's algorithms for factoring and discrete 
logarithm~\cite{Shor:factoring},
use the same generic approach sometimes called the {\em standard
method}.
The standard method for the HSP can be described as follows:
evaluate the
function $f$ in superposition and ignore the function value 
to get a state of the form
$\sigma_H := \frac{1}{|G|} \sum_{g \in G} \ketbra{gH}$, where
$\ket{gH} := \frac{1}{\sqrt{|H|}} \sum_{h \in H} \ket{gh}$, i\,.e.,
$\sigma_H$ is a uniform mixture of uniform superpositions over
left cosets $gH$ of the hidden subgroup $H$. 
A state of the form $\sigma_H$ for some subgroup $H \leq G$
is called a {\em coset state}. 
The above procedure can be repeated $t$ times to get $t$
independent copies of the state $\sigma_H$.
The aim now is to identify $H$ from $\sigma_H^{\otimes t}$.

The coset state based approach to the HSP leads us to consider the
following general problem called {\em quantum state identification}.
Given $\sigma_i^{\otimes t}$ from an a priori known ensemble 
$\cE = \{\sigma_1, \ldots, \sigma_m\}$ of quantum states in $\C^n$,
identify $i$. 
A related problem is the following {\em quantum state distinction} 
problem:
is there a single measurement basis or more generally 
a POVM $\cM$,
that gives reasonably large total variation distance between 
every pair of states in $\cE$? 
The important point here is that we want a single measurement $\cM$
that works well for every pair of states.
A solution to the state identification
problem trivially gives a solution to the state distinction problem.
It is not hard to see that the converse is also true: 
a POVM $\cM$ with {\em distinguishing power} $\delta$,
i.\.e., $\cM$ solves the state
distinction problem with total variation distance at least $\delta$
between every pair of states from $\cE$,
gives an algorithm that identifies
the given state with constant probability from 
$t = O\left(\frac{\log m}{\delta^2}\right)$ independent copies. 
This algorithm is in fact a {\em single register} algorithm in that
it applies $t$ independent copies of $\cM$ to the given 
$\sigma_i^{\otimes t}$ and does a classical `minimum-finding style'
post-processing on the observed outcomes to guess $i$.
Single register algorithms may have advantages over multi-register
algorithms in the interests of efficiency and ease of design; 
observe that the complexity of a generic $k$-register measurement
increases exponentially with $k$. 

In this work, we study information-theoretic aspects of the
general state distinction problem, and use it as a tool
for solving the corresponding state identification problem.
We also analyse various implications of these two problems, including
consequences for the HSP.
Our main objective is to
find sufficient conditions on the
ensemble $\cE$ to guarantee the existence of a 
measurement with distinguishing power $\delta$.
It is known that two quantum states can be $\delta$-distinguished 
by a measurement
if and only if they have trace distance at least $\delta$.
In general, this measurement depends upon the pair of
states to be distinguished.
Thus, this result does not give us any way to come up with a single
measurement $\cM$ is that works well for every pair of states.
However, it does provide a necessary condition:
in order for a POVM with distinguishing power $\delta$
to exist, every pair of states 
in $\cE$ must have
trace distance at least $\delta$. On a concrete note, we show that 
the ensemble of coset states for subgroups of a group $G$
indeed has minimum pairwise trace distance of $1$. 
However, constant pairwise trace 
distance is not sufficient
for the existence of a polynomially distinguishing measurement, as 
seen for example by the
recent work of Moore, Russell and Schulman~\cite{MRS:sn} on hidden 
subgroups of the symmetric group.

\paragraph{Random POVM and Frobenius distance:}
In this paper, we present for the first time a sufficient criterion 
for the state distinction
problem. Let $\frob{A}$ denote the Frobenius norm of a matrix $A$,
i.\,e.,  $\frob{A} := \sqrt{\sum_{kl} |A_{kl}|^2}$.
For a POVM $\cM$ and quantum state $\sigma$ in $\C^n$,
let $\cM(\sigma)$ denote the probability distribution on
the outcomes of $\cM$ got by measuring $\sigma$ according to
$\cM$. Our main result can be stated informally as follows.
\begin{result}[{\bf Informal statement}]
\label{res:main}
Suppose $\sigma_1$, $\sigma_2$ are two quantum states in 
$\C^n$. Define $f := \frob{\sigma_1 - \sigma_2}$. If
$\rank{\sigma_1} + \rank{\sigma_2}$ is not `too large',
then with probability at least
$1 - \exp(-\Omega(\sqrt{n}) - \exp(-\Omega(f^2 n))$ over the
choice of a random orthonormal basis $\cB$ in $\C^n$,
$\totvar{\cB(\sigma_1) - \cB(\sigma_2)} > cf$, where $c$ is
a universal constant.
\end{result}
Using the above result,
we can show that if the minimum pairwise Frobenius distance of an
ensemble $\cE = \{\sigma_1, \ldots, \sigma_m\}$ of states in $\C^n$
is at least $f$, then with probability at least
$1 - \exp(-n)$,
a random POVM $\cF$, with an appropriate notion of randomness, 
gives total variation distance at 
least $c f$ between every pair of states of $\cE$, where
$c > 0$ is a universal constant.
The notion of random POVM that we use 
is as follows: attach a zero ancilla in $\C^m$, where
$m := \Theta\left(\frac{n \log^2 m}{f^2}\right)$, and
measure 
$\sigma_i \otimes \ketbra{0}$ according to a random orthonormal
basis in $\C^n \otimes \C^m$. 
In addition, as suggested by Result~\ref{res:main},
if the maximum rank of a state in $\cE$ is not
too large, then we don't need a POVM at all, 
a random orthonormal basis in $\C^n$ will work just as well. 
We also construct examples of density matrices $\sigma_1$, $\sigma_2$
with $\trnorm{\sigma_1 - \sigma_2} = 2$,
where with very high probability the total 
variation distance
given by a random POVM is at most $\sqrt{\frob{\sigma_1 - \sigma_2}}$,
unless exponentially many ancilla qubits are used to define the 
random POVM.

\paragraph{Application to the HSP:}
Our random POVM method
has information-theoretic implications about the HSP in
a general group $G$.
It is easy to see that the
ensemble of coset states for subgroups of $G$
is simultaneously block diagonal in the
Fourier basis for $G$, where a block is labelled by 
an irreducible representation (irrep) of $G$ 
and a row index. This leads us to consider the so-called
{\em random Fourier method} for the HSP: 
apply the quantum Fourier transform over $G$ to the given coset
state and observe the name of an irrep
$\rho$ and a row index $i$, and then measure the resulting
reduced state using a random POVM.
Previously, a few examples of HSP's were given where random 
Fourier sampling required exponentially many copies of the
coset state in order to identify the hidden subgroup with
constant probability~\cite{GSVV:randbasis, MRRS:affine}.
In these examples, the ranks of the blocks of the coset state in
the Fourier basis were exponentially large.
Using the fact that 
$\frob{A} \geq \frac{\trnorm{A}}{\sqrt{\rank{A}}}$
for any matrix $A$, we
prove a surprising positive counterpart to the above
negative results. We show that
polynomially many iterations
of the random Fourier method give enough classical information
to identify the hidden subgroup $H$ if the ranks of the coset
state in each block in the Fourier basis are polynomially bounded.
In fact, we define a distance metric $r(H_1, H_2)$
between two subgroups $H_1, H_2 \leq G$ based on the Frobenius
distance between the corresponding blocks 
of the coset states $\sigma_{H_1}$ and $\sigma_{H_2}$
in the Fourier basis of $G$, and show that random Fourier
sampling gives total variation distance at least 
$\Omega(r(H_1, H_2))$ between $\sigma_{H_1}$ and $\sigma_{H_2}$
with exponentially high probability. If the ranks of the blocks
of $\sigma_{H_1}$, $\sigma_{H_2}$ are polynomially bounded,
then $r(H_1, H_2)$ is at least polynomially large.
The previous work of \cite{RRS:heisenberg} also proposed a
distance function $r'(H_1, H_2)$, but it was difficult to
estimate $r'(H_1, H_2)$ except for very special cases. Also,
the function $r'(H_1, H_2)$ is not powerful enough to even show that 
if the ranks of the blocks are $\sigma_{H_1}$, $\sigma_{H_2}$
are at most one, polynomially many iterations of random 
Fourier sampling suffice to identify the hidden subgroup with
high probability. 
Our new result improves our understanding of the power of
single register Fourier sampling, and establishes that the
random POVM method can often be a powerful information-theoretic
tool.

In particular, for the important special case when
the hidden subgroup $H$ forms a Gel'fand pair with the ambient 
group $G$, i.\,e., each block has rank either zero or one,
$O(\log^3 |G|)$ iterations of random strong Fourier sampling give 
enough classical information
to identify the hidden subgroup $H$ with high probability.
For many concrete examples e.g. affine group, Heisenberg group, 
the number of iterations of random Fourier sampling can be brought
down to $O(\log |G|)$ by a more careful analysis. Gel'fand
pairs have been studied extensively in group theory, and a lot of
recent work~\cite{MR:gelfand} on 
the hidden subgroup problem has involved Gel'fand
pairs e.g. dihedral group~\cite{EH:dihedral, BCvD:dihedral} , 
affine group~\cite{MRRS:affine}, 
Heisenberg group~\cite{RRS:heisenberg, BCvD:heisenberg}. 
For the dihedral and affine groups, it is possible to give 
explicit efficient measurement
bases for the single register
Fourier sampling procedure that identify the
hidden subgroup with high probability using polynomially many
copies. Interestingly, for the Heisenberg
group no such explicit basis for single register Fourier sampling is
known, though an explicit efficient entangled basis for
{\em two-register} Fourier sampling is 
known~\cite{BCvD:heisenberg}.
The only proof that polynomially many iterations of single 
register Fourier sampling suffice information-theoretically to 
identify hidden subgroups in the Heisenberg group is through 
random Fourier sampling, and was first observed in 
\cite{RRS:heisenberg}.

Since it can be shown that measuring in a Haar-random orthonormal
basis is
hard for a quantum computer, the main open question that
arises from our work is whether there are
efficiently implementable pseudo-random orthonormal bases for
specific ensembles that have good distinguishing power. For
example, such
a basis for the representations of groups 
$\Z_p^r \rtimes \Z_p$, $p$ prime, will give us algorithms
for the HSP in those groups having an efficient quantum part
followed by a possibly super polynomial classical post-processing.
For super constant $r$, no such quantum algorithm is currently known.
Current proposals of pseudo-random orthonormal
bases~\cite{EWSLC:pseudo, ELL:pseudo}
however, seem inadequate for our purposes.

\paragraph{Application to general state identification:}
Besides applications to the HSP, our random POVM method
also has some interesting
consequences for the general state identification problem.
For an ensemble $\cE$ of states in $\C^n$
with minimum pairwise trace distance $\delta$
and maximum rank $r$ of a state, 
$t = O\left(\frac{r \log |\cE|}{\delta^2}\right)$
independent copies of a state are enough to identify the state
with high probability using $t$ iterations of a random POVM.
Since $r \leq n$, for a general 
ensemble of quantum states we get 
$t = O\left(\frac{n \log m}{\delta^2}\right)$ which is
the first upper bound on the number of copies required
for the general state identification problem
to the best of our knowledge.
For pure states, we get 
$t = O\left(\frac{\log m}{\delta^2}\right)$ which is
optimal up to constant factors. 
This result for pure states can 
be independently
proved by a detailed analysis of Gram-Schmidt orthonormalisation,
but the resulting measurement is a {\em joint}
measurement entangled across $t$ registers.
In contrast, note that all
the state
identification algorithms arising from our random POVM result
are single register algorithms. 

\paragraph{Related work:} 
The so-called {\em pretty good measurement}, also known as the
square-root measurement, has been proposed in the past as a 
measurement for the state identification problem~\cite{HW:pgm}.
Its performance is indeed `pretty good' if the ensemble of states
possesses some special symmetries; see e.g.~\cite{EMV:pgm} and
the references therein. 
The PGM approach has been recently applied to
a few instances of
the HSP also~\cite{BCvD:dihedral, BCvD:heisenberg, MR:gelfand},
showing that it maximises the probability of identifying the
hidden subgroup for those instances. 
The PGM approach
to state identification differs from our approach in an important
way: the PGM approach does not usually give single register algorithms
for state identification, whereas our approach based on
state distinction does. This is because the PGM for $t$ copies,
in general, is a joint measurement and does not decompose as
a tensor product of measurements on the individual copies. In fact,
for the dihedral HSP studied in \cite{BCvD:dihedral}, 
an exponential number of iterations of the PGM for a single copy
are required in order to identify a hidden reflection with
constant probability. In contrast, polynomially many iterations
of `forgetful' Fourier sampling on single copies give enough
classical information to identify a hidden reflection in the
dihedral group~\cite{EH:dihedral}. 

Another problem similar to state distinction is as follows:
for two a priori known ensembles $\cE_1$, $\cE_2$ of quantum 
states, is there
a two-outcome POVM that identifies with reasonable probability 
to which ensemble a given
state from $\cE_1 \cup \cE_2$ belongs? It turns out that
the probability of error is related to the minimum trace distance 
between the convex hulls of $\cE_1$ and 
$\cE_2$~\cite{GW:convexhull, Jain:convexhull}, and is $1/2$ if the
convex hulls intersect. In contrast, in the state distinction 
problem we want to find a POVM with many outcomes that
gives reasonable total variation distance between every pair of
states of the ensemble. Having more than two outcomes allows
us to find a pairwise distinguishing POVM 
even if the ensemble cannot be partitioned
into two parts with disjoint convex hulls. 

\paragraph{Proof technique:}
In order to show that, under suitable conditions,
a random orthonormal basis $\cB$ gives
total variation distance at least 
$\Omega(\frob{\sigma_1 - \sigma_2})$ between two quantum
states $\sigma_1$, $\sigma_2$,
we have to analyse $\cB$ in the eigenbasis of $\sigma_1 - \sigma_2$.
Our techniques differ from 
earlier work on the power of random
basis for state distinction~\cite{RRS:heisenberg} in two different
ways. First, the paper \cite{RRS:heisenberg} could not handle
an arbitrary pair of quantum states $\sigma_1$, $\sigma_2$ because
of using weaker symmetry arguments. Using better symmetry arguments
and a new probabilistic
analysis of the Gram-Schmidt orthonormalisation process, we overcome
this limitation and reduce
the problem to proving lower bounds on the tail of
weighted sums of squares
of Gaussian random variables. For the pairs of states 
considered in \cite{RRS:heisenberg}, one only needed to prove
tail lower bounds for an unweighted sum of squares of Gaussian,
i.\,e., one needed to prove tail lower bounds for the chi-square
distribution. The paper \cite{RRS:heisenberg} proved such bounds
using the central limit theorem from probability theory. However,
since we are now in the weighted case, the statement of the
central limit theorem does not quite suffice. The main problem 
is that the central limit theorem cannot guarantee that
a weighted sum of squared Gaussians exceeds its mean by a
standard deviation with constant probability independent of the
number of random variables and the weights. To do this, we have
to use a powerful 
quantitative version of the central limit theorem known as the
Berry-Ess\'{e}en theorem combined with `weight smoothening'
arguments. This allows us to
show that the tail of a weighted sum of squared Gaussian
exceeds the $\ell_2$-norm of the weight vector with constant
probability. This is in
contrast to Chernoff-like upper bounds on the tail of chi-square
distributions that are more commonly seen in the study of measure
concentration for random unitaries. Since the
$\ell_2$-norm of the weight vector is closely related to
$\frob{\sigma_1 - \sigma_2}$, we get our main result easily after
this. The Berry-Ess\'{e}en theorem
also indicates that a random orthonormal basis cannot
achieve total variation distance much larger than
$\frob{\sigma_1 - \sigma_2}$, and in fact, we give an example
of states $\sigma_1$, $\sigma_2$ with trace distance $2$ where 
a random basis cannot
give total variation distance more than 
$\sqrt{\frob{\sigma_1 - \sigma_2}}$ with high probability.

\section{Preliminaries}

\subsection{Measure concentration in $\C^n$}
In this subsection, we prove some simple results about 
measure concentration phenomena in $\C^n$ for large $n$,
that will be useful in the proof of our main theorem. 

By a Gaussian probability distribution $\cG$, we mean the 
one-dimensional real Gaussian probability distribution
with mean $0$ and variance $1$, i.\,e., for $x \in \R$, the
probability density of $\cG$ at $x$ is 
$\frac{e^{-x^2/2}}{\sqrt{2 \pi}}$. 
We use $\Phi(\cdot)$ to denote the cumulative distribution function 
of $\cG$, i\,.e., $\Phi(x)$ is the probability that 
$\cG$ picks a real number less than or equal to $x$. 

The following tail bound on the sum of squares of $n$ independent
Gaussians, also known as the chi-square distribution with $n$ degrees
of freedom, can be proved Chernoff-style using the
moment generating function of the square of a Gaussian random
variable.
\begin{fact}
\label{fact:chernoffchisquare}
Let $G_1, \ldots, G_n$ be independent random variables where
each $G_i$ is distributed according to $\cG$.
Let $Y := \sum_{i=1}^n G_i^2$.
For all $\epsilon \geq 0$, 
\[
\Pr[Y > n (1 + \epsilon)] 
< (\exp(-\epsilon/2) \cdot \sqrt{1 + \epsilon})^n.
\]
The same upper bound also holds for
$\Pr[Y < n (1 + \epsilon)]$ when $-1 < \epsilon < 0$.
\end{fact}

Using Fact~\ref{fact:chernoffchisquare}, 
we can prove the following
lemma upper bounding the length of the projection
of a random unit vector onto a fixed subspace.
\begin{lemma}
\label{lem:proj}
Let $W$ be a $k$-dimensional subspace of $\C^n$, where
$k \leq n / 4$. Let $v$ be a 
random unit vector in $\C^n$. Let $\Pi_W$ denote the orthonormal
projector from $\C^n$ to $W$. Suppose $4 \leq t \leq n/k$. Then,
\[
\Pr\left[\norm{\Pi_W(v)}^2 > t \cdot \frac{k}{n}\right]
\leq \exp(-\Omega(tk)).
\]
Also, for any $0 \leq \epsilon \leq 1/2$, 
\[
\Pr
\left[
(1 - \epsilon) \frac{k}{n} \leq
\norm{\Pi_W(v)}^2 \leq 
(1 + \epsilon) \frac{k}{n}
\right]
\geq 1 - \exp(-\Omega(\epsilon^2 k)).
\]
\end{lemma}
\begin{proof}
We can choose a random
unit vector $v \in \C^n$ as
follows: choose a random vector $\hv \in \C^n$ by choosing
$2 n$ independent real random variables $G_1, \ldots, G_{2n}$,
where each $G_i$ is distributed according to $\cG$, and treating
a complex number as a pair of real numbers. Now normalise
$\hv$ to get a random unit vector $v$; note that $\norm{\hv} = 0$
with probability $0$.
By symmetry, we can assume that $W$ is spanned by the first
$k$ standard basis vectors in $\C^n$. Thus,
$
\norm{\Pi_W(v)}^2 = \frac{\sum_{i=1}^{2k} G_i^2}
                         {\sum_{j=1}^{2n} G_j^2}.
$
Using $\epsilon = -1/2$ in Fact~\ref{fact:chernoffchisquare}, we
get $\sum_{j=1}^{2n} G_j^2 > n$ with probability
at least $1 - \exp(-\Omega(n))$ over the choice of $v$.
Since 
$\exp(-\epsilon/2) \cdot \sqrt{1 + \epsilon} \leq \exp(-\epsilon/10)$
for $\epsilon \geq 1$,
using $\epsilon = t/4$ in Fact~\ref{fact:chernoffchisquare} we
get $\sum_{i=1}^{2k} G_i^2 \leq \frac{(t + 4) k}{2}$ with probability
at least $1 - \exp(-\Omega(t k))$ over the choice of $v$.
Thus, with probability at least 
$1 - \exp(-\Omega(t k)) - \exp(-\Omega(n))$ over the choice of
$v$, $\norm{\Pi_W(v)}^2 < \frac{(t + 4) k}{2n} \leq \frac{tk}{n}$.
This completes the proof of the first part of the lemma.

The proof of the second part of the lemma is very similar,
using the inequality
$\exp(-\epsilon / 2) \cdot \sqrt{1 + \epsilon} \leq -\epsilon^2/3$ 
for $0 \leq \epsilon \leq 1/2$. 
\end{proof}

We now prove a lemma upper bounding the perturbation induced by the
Gram-Schmidt orthonormalisation process on $r$ random independent
unit vectors in $\C^n$.
\begin{lemma}
\label{lem:gramschmidt}
Let $b'_1, \ldots, b'_r$ be a sequence of random independent 
unit vectors in $\C^n$, where $r \leq n$.
Let $\tb_1, \ldots, \tb_r$ be the 
corresponding
sequence of unit vectors got by Gram-Schmidt orthonormalising 
$b'_1, \ldots, b'_r$. Fix $M > 1$. Then with probability at least 
$1 - r \cdot \exp(-\Omega(M r))$ over the choice of 
$b'_1, \ldots, b'_r$,
\[
\trnorm{\ketbra{b'_i} - \ketbra{\tb_i}} 
 \leq O\left(\sqrt{\frac{M r}{n}}\right)
\]
for all $1 \leq i \leq r$,
\end{lemma}
\begin{proof}
For $1 \leq i \leq r$, let $\Pi_i$ denote the orthonormal projector
from $\C^n$ to the subspace spanned by $b'_1, \ldots, b'_i$.
For $1 \leq i \leq r - 1$, putting $t = \frac{M r}{i}$ in 
the first part of Lemma~\ref{lem:proj}, we see that 
with probability at least 
$1 - r \exp(-\Omega(M r))$ over the
choice of $b'_1, \ldots, b'_r$,
$\norm{\Pi_i(b'_{i+1})}^2 \leq O\left(\frac{M r}{n}\right)$. 
Recall that 
$
\tb_{i+1} := \frac{b'_{i+1} - \Pi_i(b'_{i+1})}
                  {\norm{b'_{i+1} - \Pi_i(b'_{i+1})}}.
$
Hence, 
\begin{eqnarray*}
&   &
\!\!\!\!\!
\!\!\!\!\!
\!\!
\norm{\tb_{i+1} - b'_{i+1}}^2 
    =    \norm{\Pi_i(b'_{i+1})}^2 + 
         \left(1 - \norm{b'_{i+1} - \Pi_i(b'_{i+1})}\right)^2 \\
&   =  & \norm{\Pi_i(b'_{i+1})}^2 +
         \left(1 - \sqrt{1 - \norm{\Pi_i(b'_{i+1})}^2}\right)^2 
    =    2 - 2 \sqrt{1 - \norm{\Pi_i(b'_{i+1})}^2} \\
& \leq & 2 - 2 \sqrt{1 - O\left(\frac{M r}{n}\right)}
  \leq   O\left(\frac{M r}{n}\right).
\end{eqnarray*}

The proposition now follows from the fact that for two unit vectors
$\ket{\psi}$, $\ket{\phi}$, 
$
\trnorm{\ketbra{\psi} - \ketbra{\phi}} 
\leq 2 \norm{\ket{\psi} - \ket{\phi}}.
$
\end{proof}

We will require the following fact about the size of
a $\delta$-net in $\C^n$. A $\delta$-net $\cN$ is a finite set
of unit vectors in $\C^n$ with the property that for any unit
vector $v \in \C^n$, there exists a unit vector $v' \in \cN$
such that $\norm{v - v'} \leq \delta$. The fact follows
from the proof technique of 
\cite[Lemma~13.1.1, Chapter~13]{Matousek:dg} and by identifying
$\C^n$ with $\R^{2n}$.
Below for $1 \leq j \leq n$, $e_j$ denotes the $j$th standard
unit vector in $\C^n$, viz., the $n$-tuple containing a $1$ in
the $j$th location and zeroes elsewhere.
\begin{fact}
\label{fact:deltanet}
Fix any $\delta \in (0, 1]$. Then, there is a $\delta$-net $\cN$ in
$\C^n$ containing the $n$ standard unit vectors $e_1, \ldots, e_n$
such that $|\cN| \leq \left(\frac{4}{\delta}\right)^{2n}$.
\end{fact}

Using Fact~\ref{fact:deltanet}, we can prove the following
lemma upper bounding the spectral norm of an
$n \times n$ matrix whose entries are independent random complex
numbers with independent Gaussian real and imaginary parts.
\begin{lemma}
\label{lem:normrandmatrix}
Define a random $n \times n$ complex matrix $M$ by independently
choosing each entry to be a complex number whose real and imaginary
parts are independently chosen according to the Gaussian distribution
$\cG$. Then, with probability at least $1 - \exp(-\Omega(n \log n))$
over the choice of $M$, $\norm{M} \leq O(\sqrt{n \log n})$.
\end{lemma}
\begin{proof}
Let $\delta := 1/\sqrt{n}$. 
Let $\cN$ be a $\delta$-net in $\C^n$ guaranteed by 
Fact~\ref{fact:deltanet}.
Fix any unit vector $v \in \C^n$. By symmetry, the probability
distribution of $\norm{M v}^2$ is the same as that of
$\norm{M e_1}^2$, i.\,e., the probability distribution of
$\norm{M v}^2$ is the same as that of the sum of squares
of $2 n$ independent
Gaussians. Let $t := C \log n$, where $C$ is a sufficiently
large constant whose value will become clear later. Since 
$\exp(-\epsilon/2) \cdot \sqrt{1 + \epsilon} \leq \exp(-\epsilon/10)$
for $\epsilon \geq 1$, using $\epsilon = t$ in
Fact~\ref{fact:chernoffchisquare}, we get that
$\norm{M v'}^2 \leq (t + 1) n$ for all $v' \in \cN$
with probability at least
$
1 - (4\sqrt{n})^{2n} \cdot \exp(-\Omega(C n \log n))
\geq 1 - \exp(-\Omega(n \log n))
$ 
over the choice of $M$.

Note that for any vector $w \in \C^n$, we have
\begin{eqnarray*}
&   &
\!\!\!\!\!
\!\!\!\!\!
\!\!
\norm{M w}^2 
    =    \sum_{i=1}^n \left|\sum_{j=1}^n M_{ij} w_j\right|^2
  \leq   \sum_{i=1}^n \left(\sum_{j=1}^n |M_{ij}|^2\right) \cdot
                      \left(\sum_{j=1}^n |w_j|^2\right)
    =    \norm{w}^2 \sum_{j=1}^n \sum_{j=1}^n |M_{ij}|^2 \\
&   =  & \norm{w}^2 \sum_{j=1}^n \norm{M e_j}^2
  \leq   \norm{w}^2 n^2 (t + 1).
\end{eqnarray*}
The inequality above follows from Cauchy-Schwartz.
Now fix any unit vector $v \in \C^n$. Let $v'$ be the closest
vector to $v$ from $\cN$, where ties are broken arbitrarily.
Thus, $\norm{v - v'} \leq \delta$.
We have
\begin{eqnarray*}
&   &
\!\!\!\!\!
\!\!\!\!\!
\!\!
\norm{M v}^2 
    =    \qform{v}{M^\dag M}{v}
    =    \qform{v' + (v - v')}{M^\dag M}{v' + (v - v')} \\
&   =  & \norm{M v'}^2 + \qform{v'}{M^\dag M}{v - v'} +
         \qform{v - v'}{M^\dag M}{v'} + \norm{M (v - v')}^2 \\
& \leq & \norm{M v'}^2 + 2 \norm{M v'}\norm{M (v - v')} +
         \norm{M (v - v')}^2 \\
& \leq & (t + 1)n + 
         2 \sqrt{(t+1)n} \cdot \norm{v - v'} \cdot n \sqrt{t+1} + 
         \norm{v - v'}^2 n^2 (t + 1) \\
& \leq & (t + 1)n + 
         2 n^{3/2} (t+1) \delta + \delta^2 n^2 (t+1)
  \leq   O(n \log n).
\end{eqnarray*}
The first
inequality above follows from Cauchy-Schwartz. The proof of
the lemma is now complete.
\end{proof}

Finally, we will require the 
following Berry-Ess\'{e}en theorem from 
probability theory, which is a quantitative version of the
central limit 
theorem~\cite[Chapter~XVI, Section~5, Theorem~2]{feller:vol2}.
\begin{fact}[{\bf Berry-Ess\'{e}en theorem}]
\label{fact:berryesseen}
Let $X_1, \ldots, X_n$ be independent random variables.
Define $\mu_i := \E[X_i]$, $\sigma_i := (\E[|X_i - \mu_i|^2])^{1/2}$,
$\rho_i := (\E[|X_i - \mu_i|^3])^{1/3}$. Define the quantities
\[
\sigma^2 := \sum_{i=1}^n \sigma_i^2,
\qquad
\rho^3 := \sum_{i=1}^n \rho_i^3,
\qquad
X := \frac{1}{\sigma} \sum_{i=1}^n (X_i - \mu_i).
\]
Then for all $x \in \R$,
\[
|\Pr[X \leq x] - \Phi(x)| \leq \frac{6 \rho^3}{\sigma^2}.
\]
\end{fact}
\paragraph{Remark:}
The constant $6$ in the Berry-Ess\'{e}en theorem can be improved;
the current record is $0.7915$ by 
Shiganov~\cite{Shiganov:berryesseen}. However,
Proposition~\ref{prop:berryesseen} below
holds as long as the constant is finite and independent of $n$ 
and the random variables $X_1, \ldots, X_n$.  

Using Fact~\ref{fact:berryesseen}, we 
prove the following proposition which will play a central role
in the proof of our main theorem. 
\begin{proposition}
\label{prop:berryesseen}
Let $G_1, \ldots, G_n$ be independent random variables where
each $G_i$ is distributed according to $\cG$.
Let $\lambda_1, \ldots, \lambda_n \in (0, 1]$.
Define 
\[
t := \sum_{i=1}^n \lambda_i,
\qquad
f := \sqrt{\sum_{i=1}^n \lambda_i^2},
\qquad
X := \sum_{i=1}^n \lambda_i G_i^2.
\]
Suppose $t \leq 1$.
Then, there is a
constant $c$ independent of $n$ and $\lambda_1, \ldots, \lambda_n$
such that 
\[
\Pr[X > t + f] > c 
\qquad
\mbox{{\rm and}}
\qquad
\Pr[X < t] > c.
\]
\end{proposition}
\begin{proof}
Without loss of generality, $\lambda_1 \geq \cdots \geq \lambda_n$.
Let $K_1$ be a sufficiently large constant, whose choice will become
clear later.
Suppose $\lambda_1 \geq \frac{t}{K_1}$. 
Note that $\frac{t}{K_1} \leq f \leq t$.
There is a constant $c_1$
depending on $K_1$ but
independent of $n$ and $\lambda_1, \ldots, \lambda_n$
such that $\Pr[G_1^2 > 2 K_1] > c_1$, which implies that
\[
\Pr[X > t + f] > \Pr[\lambda_1 G_1^2 > 2 t] 
               > \Pr\left[\frac{t}{K_1} G_1^2 > 2 t\right]
               = \Pr\left[G_1^2 > 2 K_1\right]
               > c_1.
\]
Also, 
\begin{eqnarray*}
&   &
\!\!\!\!\!
\!\!\!\!\!
\!\!\!\!\!
t 
        =       \E[X] 
      \geq      t \cdot \Pr[t \leq X \leq t + f] + 
                (t + f) \Pr[X > t + f] \\
&       =     & t \cdot \Pr[X \geq t] + f \cdot \Pr[X > t + f] \\
&     \geq    & t \cdot \Pr[X \geq t] + \frac{t}{K_1} \cdot c_1 \\
&       =     & t \cdot (1 - \Pr[X < t]) + \frac{t c_1}{K_1} \\
& \Rightarrow & \Pr[X < t] \geq \frac{c_1}{K_1}.
\end{eqnarray*}
Now, suppose $\lambda_1 < \frac{t}{K_1}$. Define independent random
variables $X_i := \lambda_i G_i^2$. 
Let $\mu_i$, $\sigma_i$, $\rho_i$ be defined as in 
Fact~\ref{fact:berryesseen}.
Recall that $\E[G_i^2] = 1$, $\E[|G_i^2 - 1|^2] = 2$ and that
the absolute third central moment of $G_i^2$ is finite, say equal
to $K_2$. Then,
\[
\frac{6 \sum_{i=1}^n \rho_i^3}{\sum_{i=1}^n \sigma_i^2} 
  =  \frac{6 K_2 \sum_{i=1}^n \lambda_i^3}
          {2 \sum_{i=1}^n \lambda_i^2} 
  <  \frac{6 K_2 t}{2 K_1}
\leq \frac{3 K_2}{K_1}.
\]
Taking $x = \frac{1}{\sqrt{2}}$ in Fact~\ref{fact:berryesseen}, 
we get
\[
\Pr[X > t + f] 
\geq \left(1 - \Phi\left(\frac{1}{\sqrt{2}}\right)\right)
     - \frac{3 K_2}{K_1}.
\]
Similarly, 
taking $x = 0$ in Fact~\ref{fact:berryesseen} we get
\[
\Pr[X \leq t] 
\geq \Phi(0) - \frac{3 K_2}{K_1}
  =  \frac{1}{2} - \frac{3 K_2}{K_1}.
\]
Choosing $K_1$ to be a sufficiently large constant, we see that
there exists a universal constant $c_2$ such that
$\Pr[X > t + f] > c_2$ and $\Pr[X < t] = \Pr[X \leq t] > c_2$.
Now letting $c := \min\left\{\frac{c_1}{K_1}, c_2\right\}$,
we have that $\Pr[X > t + f] > c$ and $\Pr[X < t] > c$ always. 
Observe that 
$c$ is a universal constant independent of $n$ and
$\lambda_1, \ldots, \lambda_n$.
\end{proof}

\subsection{Quantum state distinction versus identification}
In this subsection, we explore the connection between the problems
of quantum state distinction and state identification.

A quantum state in $\C^n$ is modelled by a {\em density matrix}
$\sigma$, which is an $n \times n$ Hermitian, positive semidefinite
matrix with unit trace. A {\em positive operator-valued measure},
or POVM for short, is
the most general measurement on 
quantum states. See e.g.~\cite{NC:quantumbook} for a good
introduction to density matrices and POVM's. 
A POVM $\cM$ in $\C^n$ is a finite collection of
positive operators $E_i$ on $\C^n$, called elements of $\cM$, that
satisfy the completeness
condition $\sum_i E_i = \onemat_n$. If the state of the quantum system
is given by the density matrix $\sigma$, then the probability $p_i$ to
observe outcome labelled $i$ is given by the Born rule 
$p_i = \Tr(\sigma E_i)$. 
We use $\cM(\sigma)$ to denote the probability distribution
on the outcomes of $\cM$ got by measuring $\sigma$ according to 
$\cM$. The {\em trace norm} of an $n \times n$ matrix 
$A$ is defined as
$\trnorm{A} := \Tr \sqrt{A^\dag A}$. 
The {\em Frobenius norm} of $A$ is defined as  
$\frob{A} := \sqrt{\Tr A^\dag A}$, which is nothing but the 
$\ell_2$-norm of the long vector in $\C^{n^2}$ corresponding 
to $A$. The following fact follows easily from the Cauchy-Schwartz
inequality.
\begin{fact}
\label{fact:trnormfrobnorm}
For any matrix $A$, 
$\frob{A} \geq \frac{\trnorm{A}}{\sqrt{\rank{A}}}$.
\end{fact}

Suppose there is an a priori known ensemble 
$\cE = \{\sigma_1, \ldots, \sigma_m\}$ of quantum states in $\C^n$.
Given $t$ copies of a state $\sigma_i$,
a {\em single register state identification} algorithm $\cA$
for the ensemble $\cE$
consists of a sequence of POVM's $\cF_j$, $1 \leq j \leq t$, where 
$\cF_j$ operates on the $j$th copy of $\sigma_i$. There is no
bound on the number of outcomes of $\cF_j$. 
The choice of
$\cF_j$ may depend on the observed outcomes of 
$\cF_1, \ldots, \cF_{j-1}$. After $t$ observations, $\cA$ does
a classical post-processing and declares its guess
for $i$. For all $i$, $1 \leq i \leq m$, we want $\cA$ to guess
$i$ with probability at least $3/4$.

Let $0 \leq \delta \leq 2$.
A POVM $\cM$ for the {\em state distinction} problem with
{\em distinguishing power} $\delta$ for the ensemble $\cE$
is a POVM with the property
that $\totvar{\cM(\sigma_i) - \cM(\sigma_j)} \leq \delta$ for
all $1 \leq i < j \leq m$. It is easy to see via the triangle
inequality that if there exists a single register state
identification POVM on $t$ copies, then there exists a 
state distinction POVM with distinguishing power $\Omega(1/t)$.
The following fact is a converse to the above observation;
a proof sketch is included for completeness.
\begin{fact}
\label{fact:minfinding}
Let $\cE = \{\sigma_1, \ldots, \sigma_m\}$ be 
an a priori known ensemble of quantum states in $\C^n$.
If there is a POVM $\cM$ for the state distinction problem with
distinguishing power $\delta$ for the ensemble $\cE$,
then there is a single register state identification algorithm
$\cA$ for ensemble $\cE$ working on 
$t = O\left(\frac{\log m}{\delta^2}\right)$.
\end{fact}
\begin{proof}
Fix $1 \leq i < j \leq m$.
Under the promise that the unknown state is
either $\sigma_i$ or $\sigma_j$, applying $\cM$ to each of
$t$ copies of the unknown state followed by a maximum
likelihood estimate identifies the correct state
with probability at least $1 - \frac{1}{4m}$, as can be seen
by a standard Chernoff bound. Let $F_{ij}$ denote this maximum
likelihood routine.
The identification algorithm $\cA$ starts by applying 
$\cM$ on each of $t$ copies of the unknown state, which a priori
can be any $\sigma_i \in \cE$. 
After that, $\cA$ does $m-1$ iterations of a
classical minimum-finding style post-processing procedure comparing
two possible states $\sigma_i$, $\sigma_j$ in an iteration,
using the classical routines $F_{ij}$
on the $t$ observed outcomes. Note that the same $t$ observed
outcomes are reused by the various routines $F_{ij}$; no fresh
measurements are done. The success probability of the 
minimum-finding style post-processing, and hence algorithm $\cA$, 
is at least $1 - \frac{m-1}{4m}  \geq 3/4$.
\end{proof}

\subsection{Hidden subgroup problem and quantum Fourier transform}
\label{subsec:hspqft}
In this section, we explain the importance of the quantum 
Fourier transform
as a means of attacking the hidden subgroup problem.
For a general introduction to representation theory of finite groups,
see e.g.~\cite{serre:reptheory}.

We use the term irrep
to denote an irreducible unitary representation of a finite group $G$
and denote by $\hG$ a complete set of inequivalent irreps.  For any
unitary representation $\rho$ of $G$, let $\rho^\ast$ denote the
representation obtained by entry-wise conjugating the unitary matrices
$\rho(g)$, where $g \in G$. Note that the definition of $\rho^\ast$
depends upon the choice of the basis used to concretely describe the
matrices $\rho(g)$. If $\rho$ is an irrep of $G$ so is $\rho^\ast$,
but in general $\rho^\ast$ may be inequivalent to $\rho$.  Let
$V_\rho$ denote the vector space of $\rho$, define $d_\rho := \dim
V_\rho$, and notice that $V_\rho = V_{\rho^\ast}$.  The group elements
$\ket{g}$, where $g \in G$ form an orthonormal basis of $\C^{|G|}$.
Since $\sum_{\rho \in \hG} d_\rho^2 = |G|$, we can consider another
orthonormal basis called the {\em Fourier basis} of $\C^{|G|}$ 
indexed
by $\ket{\rho, i, j}$, where $\rho \in \hG$ and $i, j$
run over the row and column indices of $\rho$. The quantum Fourier
transform over $G$, $\QFT_G$ is the following linear transformation:
\[
\ket{g} \mapsto \sum_{\rho\in \hG} \sqrt{\frac{d_\rho}{|G|}}
\sum_{i, j = 1}^{d_\rho} \rho_{i j}(g) \ket{\rho, i, j}.
\]
It follows from Schur's orthogonality relations 
(see e.g.~\cite[Chapter 2, Proposition 4, Corollary 3]
               {serre:reptheory})
that $\QFT_G$ is
a unitary transformation in $\C^{|G|}$.

For a subgroup $H \leq G$ and $\rho \in \hG$, 
define 
$\rho(H) := \frac{1}{|H|} \sum_{h \in H} \rho(h)$. 
It follows from Schur's lemma 
(see e.g.~\cite[Chapter 2, Proposition 4]{serre:reptheory})
that $\rho(H)$ is an
orthogonal projection to the subspace of $V_\rho$ consisting of
vectors that are point-wise fixed by every $\rho(h)$, $h \in H$.
Define $r_\rho(H) := \rank{\rho(H)}$. 
Notice that $r_\rho(H) = r_{\rho^\ast}(H)$.
The {\em standard method} of attacking the HSP
in $G$ using coset states~\cite{GSVV:randbasis} starts by
forming the uniform superposition $\frac{1}{\sqrt{|G|}} \sum_{g \in G}
\ket{g}\ket{0}$. It then queries $f$ to get the superposition
$\frac{1}{\sqrt{|G|}} \sum_{g \in G} \ket{g} \ket{f(g)}$. Ignoring the
second register the reduced state on the first register becomes the
density matrix $\sigma_H = \frac{1}{|G|} \sum_{g \in G}
\ket{gH}\bra{gH}$, that is the reduced state is a uniform mixture over
all left coset states of $H$ in $G$.  It can be easily seen that
applying $\QFT_G$ to $\sigma_H$ gives us the density matrix
$\frac{|H|}{|G|} \bigoplus_{\rho \in \hG} \bigoplus_{i=1}^{d_\rho}
\ket{\rho,i}\bra{\rho,i} \otimes \rho^\ast(H)$, where $\rho^\ast(H)$
operates on the space of column indices of $\rho$. 
Since the states $\sigma_H$ are simultaneously
block diagonal in the Fourier
basis for any $H\leq G$, the elements of any POVM $\cM$
operating on these
states can without loss of generality be assumed to have the same
block structure.  From this it is clear that any 
distinguishing measurement 
without loss of generality first applies the quantum
Fourier transform $\QFT_G$ to $\sigma_H$, measures the name $\rho$
of an irrep, the index $i$ of a row, and then measures the reduced
state on the column space of $\rho$ using a POVM $\cM_\rho$
in $\C^{d_\rho}$. This POVM $\cM_\rho$ may depend on 
$\rho$ but is independent of $i$. 

The probability of observing an irrep $\rho$  in this quantum state
is given by
$\cP_H(\rho) = \frac{d_{\rho} |H| r_{\rho}(H)}{|G|}$. 
Conditioned on observing $\rho$ we obtain a
uniform distribution $1/d_\rho$ on the row indices. The reduced state
on the space of column indices
after having observed an irrep $\rho$ and a row index $i$ is then
given by the state $\rho^\ast(H)/r_\rho(H)$, and a basic task for a
hidden subgroup finding algorithm
is how to extract information about $H$ from it.
In this paper, we will investigate
the case when $\cM_\rho$ is a random POVM, for a suitable
definition of randomness, in $\C^{d_\rho}$. We shall call this
procedure {\em random Fourier sampling}. 
Grigni, Schulman, Vazirani and Vazirani~\cite{GSVV:randbasis} 
show that under certain conditions on
$G$ and $H$, random Fourier sampling gives
exponentially small information about distinguishing $H$ from the
identity subgroup.  In this paper, we prove a complementary
information-theoretic
result viz. under different conditions on $G$, $(\log |G|)^{O(1)}$
random strong Fourier samplings do give enough information to
reconstruct the hidden subgroup $H$ with high probability.

In {\em weak Fourier sampling}, we only measure the name 
of an irrep and ignore the reduced state
on the column space.
It can be shown~\cite{HRT:normalhsp} that for normal
hidden subgroups $H$, no more information about $H$ is contained 
in the reduced state.
Thus, weak Fourier sampling is the optimal measurement
to recover a normal hidden subgroup from its 
coset state. In particular, Fourier sampling is the optimal 
measurement on coset states for the abelian HSP.

Define a distance metric
$
w(H_1, H_2) := \totvar{\cP_{H_1} - \cP_{H_2}} =
\sum_{\rho \in \hG} |\cP_{H_1}(\rho) - \cP_{H_2}(\rho)|
$
between subgroups $H_1, H_2 \leq G$.
Adapting an argument in \cite{HRT:normalhsp}, it can be shown that 
$w(H_1, H_2) \geq 1/2$ if the {\em normal cores} of $H_1$ and
$H_2$ are different~\cite{RRS:heisenberg}. Recall that the
normal core of a subgroup $H$ is the
largest normal subgroup of 
$G$ contained in $H$.
Thus, the main 
challenge is to distinguish between hidden subgroups $H_1$, $H_2$ 
from the same normal core family.

We next show that coset states
corresponding to different hidden subgroups of a group have 
trace distance at least $1$. 
\begin{proposition}
\label{prop:hsptrdist}
Let $H_1$, $H_2$ be different subgroups of a group $G$. Then,
$\trnorm{\sigma_{H_1} - \sigma_{H_2}} \geq 1$.
\end{proposition}
\begin{proof}
For a subgroup $H \leq G$, we let $G/H$ denote a complete set
of left coset representatives of $H$ in $G$. Since
for any $c_1 \in G/H_1$, 
\[
\ket{c_1 H_1} = \sqrt{\frac{|H_1 \cap H_2|}{|H_1|}}
\sum_{\stackrel{c \in G/(H_1 \cap H_2)}{c H_1 = c_1 H_1}}
\ket{c (H_1 \cap H_2)},
\]
we get
\[
\sigma_{H_1} = 
\frac{|H_1|}{|G|} \sum_{c_1 \in G/H_1} \ketbra{c_1 H_1} = 
\frac{|H_1 \cap H_2|}{|G|} 
\sum_{\stackrel{c, c' \in G/(H_1 \cap H_2)}{c H_1 = c' H_1}}
\ket{c (H_1 \cap H_2)}\bra{c' (H_1 \cap H_2)}.
\]
A similar fact is true for $\sigma_{H_2}$.
We now define
\[
\hsigma_{H_1} := 
\frac{|H_1 \cap H_2|}{|G|} 
\sum_{\stackrel{c, c' \in G/(H_1 \cap H_2)}
               {c H_1 = c' H_1, c \neq c'}
     }
\ket{c (H_1 \cap H_2)}\bra{c' (H_1 \cap H_2)}.
\]
We define $\hsigma_{H_2}$ similarly.
Note that $\hsigma_{H_1}$, $\hsigma_{H_2}$ are Hermitian and
for any $c \in G/(H_1 \cap H_2)$,
$\qform{c (H_1 \cap H_2)}{\hsigma_{H_1}}{c (H_1 \cap H_2)} = 0$
and
$\qform{c (H_1 \cap H_2)}{\hsigma_{H_1}}{c (H_1 \cap H_2)} = 0$.

We now observe that for any $c, c' \in G/(H_1 \cap H_2)$,
\[
(\qform{c (H_1 \cap H_2)}{\sigma_{H_1}}{c' (H_1 \cap H_2)} \neq 0)
\wedge
(\qform{c (H_1 \cap H_2)}{\sigma_{H_2}}{c' (H_1 \cap H_2)} \neq 0)
\iff
c = c'. 
\]
This is because $c H_1 = c' H_1$ and
$c H_2 = c' H_2$ implies that $c (H_1 \cap H_2) = c' (H_1 \cap H_2)$,
i.\,e. $c = c'$. 
This implies that for any $c, c' \in G/(H_1 \cap H_2)$,
\[
(\qform{c (H_1 \cap H_2)}{\hsigma_{H_1}}{c' (H_1 \cap H_2)} = 0)
\vee
(\qform{c (H_1 \cap H_2)}{\hsigma_{H_2}}{c' (H_1 \cap H_2)} = 0).
\]
Thus, 
$
\hsigma_{H_1} \hsigma_{H_2} =
\hsigma_{H_2} \hsigma_{H_1} = 0.
$
Also, it follows that
$\sigma_{H_1} - \sigma_{H_2} = \hsigma_{H_1} - \hsigma_{H_2}$.
 
Without loss of generality, $H_1$ is not a subgroup of $H_2$. 
Now,
\begin{eqnarray*}
&      &
\!\!\!\!\!
\!\!\!\!\!
\!\!
\trnorm{\sigma_{H_1} - \sigma_{H_2}}
    =    \trnorm{\hsigma_{H_1} - \hsigma_{H_2}}
    =    \Tr \sqrt{(\hsigma_{H_1} - \hsigma_{H_2})^2} \\
&   =  & \Tr \sqrt{\hsigma_{H_1}^2 + \hsigma_{H_2}^2 
                   - \hsigma_{H_1} \hsigma_{H_2}
                   - \hsigma_{H_2} \hsigma_{H_1}
                  } \\
&   =  & \Tr \sqrt{\hsigma_{H_1}^2 + \hsigma_{H_2}^2}
  \geq   \Tr \sqrt{\hsigma_{H_1}^2}
    =    \trnorm{\hsigma_{H_1}}.
\end{eqnarray*}
The inequality follows from the fact that $\hsigma_{H_1}^2$,
$\hsigma_{H_2}^2$ are positive semidefinite operators and the 
square-root function is monotonically increasing for such operators.
In order to evaluate $\trnorm{\hsigma_{H_1}}$, notice that 
$
\hsigma_{H_1} = 
\frac{|H_1 \cap H_2|}{|G|} 
\bigoplus_{c_1 \in G/H_1} M_{c_1},
$
where for any $c_1 \in  G/H_1$,
\[
M_{c_1} :=
\sum_{\stackrel{c, c' \in G/(H_1 \cap H_2)} 
               {\stackrel{c H_1 = c' H_1 = c_1 H_1}{c \neq c'}}
     }
\ket{c (H_1 \cap H_2)}\bra{c' (H_1 \cap H_2)}.
\]
Now observe that $M_{c_1}$ is of the form $J - I$, where 
$J$, $I$ are the 
$\frac{|H_1|}{|H_1 \cap H_2|} \times \frac{|H_1|}{|H_1 \cap H_2|}$ 
all ones and identity matrices respectively.
Hence, 
$\trnorm{M_{c_1}} = 2 \left(\frac{|H_1|}{|H_1 \cap H_2|} - 1\right)$ 
for all $c_1 \in  G/H_1$.
Thus, 
\[
\trnorm{\hsigma_{H_1}} = 
\frac{|H_1 \cap H_2|}{|G|} \cdot
\frac{|G|}{|H_1|} \cdot
2 \left(\frac{|H_1|}{|H_1 \cap H_2|} - 1\right) =
\frac{2 (|H_1| - |H_1 \cap H_2|)}{|H_1|} \geq
1.
\]
The inequality follows from the fact that
$H_1 \cap H_2$ is a proper subgroup of $H_1$, since
$H_1$ is not a subgroup of $H_2$.
This completes the proof of the proposition.
\end{proof}

\section{Random measurement bases and Frobenius distance}
In this section, we prove our main result showing that 
a random POVM, for a suitable definition of randomness,
distinguishes between two density matrices by at least their
Frobenius distance with high probability.
We first prove an important technical lemma that quickly implies
our main theorem.
\begin{lemma}
\label{lem:frob}
Let $\sigma_1$, $\sigma_2$ be two density matrices in $\C^n$.
Define $f := \frob{\sigma_1 - \sigma_2}$. Then:
\begin{enumerate}
\item
If $\rank{\sigma_1} + \rank{\sigma_2} \leq \sqrt{n}/K$,
where $K$ is a sufficiently large universal constant,
then with probability at least
$
1 - \exp(-\Omega(\sqrt{n})) - 
\frac{\sqrt{n}}{K} \cdot \exp(-\Omega(f^2 n))
$
over the choice of a random orthonormal measurement basis $\hcB$ in
$\C^n$, $\totvar{\hcB(\sigma_1) - \hcB(\sigma_2)} > \Omega(f)$;

\item
Take a set $\cB$ of $n$ independent random vectors 
$\cB := \{b_1, \ldots, b_n\}$ in 
$\C^n$, where each $b_i$ is got by choosing $n$ independent
complex numbers whose real and imaginary parts are independently
chosen according to the Gaussian $\cG$.  Define
$\ell := \norm{\sum_{i=1}^n b_i b_i^\dag}$ and
$
\nu := \onemat_{\C^n} - \frac{1}{\ell} 
       \sum_{i=1}^n b_i b_i^\dag.
$
Let $\cM$ denote the POVM on $\C^n$ 
consisting of the elements 
$\frac{b_i b_i^\dag}{\ell}$ for $1 \leq i \leq n$,
and the element $\nu$. Note that $\cM$ can be implemented
as an orthonormal measurement in $\C^n \otimes \C^2$.
Then with probability at least $1 - \exp(-\Omega(n))$ over the
choice of $\cB$, 
$
\totvar{\cM(\sigma_1) - \cM(\sigma_2)} >
\Omega\left(\frac{f}{\log n}\right).
$
\end{enumerate}
\end{lemma}
\begin{proof}
We start by proving the first part of the lemma.
Define $t := \trnorm{\sigma_1 - \sigma_2}$. 
We have
$\rank{\sigma_1 - \sigma_2} \leq \sqrt{n}/K$,
where $K$ is a sufficiently large universal constant whose value will
become clear later. 
Let $\hcB := \{\ket{\hb_1}, \ldots, \ket{\hb_n}\}$ be a 
random orthonormal
basis of $\C^n$. Let $\hcB(\sigma_1)$, $\hcB(\sigma_2)$ denote 
the probability
distributions on $[n]$ got by measuring $\sigma_1$, $\sigma_2$
respectively according to
$\hcB$. Let $\lambda_1, \ldots, \lambda_k$ denote the positive
eigenvalues,
and $-\mu_{k+1}, \ldots, -\mu_{k+l}$ the negative
eigenvalues of $\sigma_1 - \sigma_2$. Note that
$
k + l   =  \rank{\sigma_1 - \sigma_2}
      \leq \sqrt{n}/K.
$
We assume that we work in the eigenbasis of $\sigma_1 - \sigma_2$.
Hence, we can write
\[
\sigma_1 - \sigma_2 = \sum_{i=1}^k \lambda_i \ketbra{i} -
                      \sum_{j=k+1}^{k+l} \mu_j \ketbra{j},
\qquad
\sum_{i=1}^k \lambda_i = \sum_{j=k+1}^{k+l} \mu_j = \frac{t}{2},
\qquad
\sum_{i=1}^k \lambda_i^2 + \sum_{j=k+1}^{k+l} \mu_j^2 = f^2.
\]
Without loss of generality, 
$
\sum_{i=1}^k \lambda_i^2 \geq \sum_{j=k+1}^{k+l} \mu_j^2
\Rightarrow \sum_{i=1}^k \lambda_i^2 \geq f^2/2.
$
Also, by the Cauchy-Schwartz inequality
$t \leq f \sqrt{k+l}$.
Then, 
\begin{eqnarray*}
&   &
\!\!\!\!\!
\!\!\!\!\!
\!\!
\totvar{\hcB(\sigma_1) - \hcB(\sigma_2)} 
  =  
\sum_{t=1}^n \left|\qform{\hb_t}{\sigma_1}{\hb_t} - 
                   \qform{\hb_t}{\sigma_2}{\hb_t}
             \right|
  =  
\sum_{t=1}^n \left|\qform{\hb_t}{\sigma_1 - \sigma_2}{\hb_t}
             \right| \\
& = &
\sum_{t=1}^n 
\left|
\sum_{i=1}^k \lambda_i \left|\braket{\hb_t}{i}\right|^2 -
\sum_{j=k+1}^{k+l} \mu_j \left|\braket{\hb_t}{j}\right|^2
\right|.
\end{eqnarray*}

Define the random $n \times n$ unitary matrix 
$\hcB$ to be the matrix whose row vectors are
$\bra{\hb_1}, \ldots, \bra{\hb_n}$. Then,
$
\totvar{\hcB(\sigma_1) - \hcB(\sigma_2)} =
\sum_{t=1}^n 
\left|
\sum_{i=1}^k \lambda_i |\hcB_{ti}|^2 -
\sum_{j=k+1}^{k+l} \mu_j |\hcB_{tj}|^2
\right|.
$
Instead of generating the random unitary matrix $\hcB$ row-wise,
we can generate it column-wise. The advantage now is that 
we only have to randomly generate the first $k + l$ orthonormal 
columns; the rest of the columns can be assumed to be zero without
loss of generality. That is, we 
generate an $n \times (k+l)$ matrix $\tcB$ whose columns are 
random orthonormal vectors $\ket{\tb_1}, \ldots, \ket{\tb_{k+l}}$
in $\C^n$. To generate the matrix $\tcB$, we generate an
$n \times (k+l)$ matrix $\cB'$ whose columns are 
random independent unit vectors $\ket{b'_1}, \ldots, \ket{b'_{k+l}}$
in $\C^n$, and apply Gram-Schmidt orthonormalisation to get
$\ket{\tb_1}, \ldots, \ket{\tb_{k+l}}$. Choosing 
$M = \frac{n}{K^2 (k+l)^2}$
in Lemma~\ref{lem:gramschmidt}, we get
$
\trnorm{\ketbra{\tb_t} - \ketbra{b'_t}} 
< O\left(\frac{1}{K \sqrt{k+l}}\right)
$ 
for all $1 \leq t \leq k + l$ with probability at least
$
1 - (k+l) \exp\left(-\Omega\left(\frac{n}{K^2 (k+l)}\right)\right) 
\geq 1 - \exp(-\Omega(\sqrt{n}/K))
$ over the choice of $\cB'$.
Let $\tcB(\sigma_1) - \tcB(\sigma_2)$ and
$\cB'(\sigma_1) - \cB'(\sigma_2)$ denote the 
functions on $[n]$ defined by 
\[
\begin{array}{l}
(\tcB(\sigma_1) - \tcB(\sigma_2))(t) 
:= \sum_{i=1}^k \lambda_i |\braket{\tb_i}{t}|^2 -
   \sum_{j=k+1}^{k+l} \mu_j |\braket{\tb_j}{t}|^2 
 = \sum_{i=1}^k \lambda_i |\tcB_{ti}|^2 -
   \sum_{j=k+1}^{k+l} \mu_j |\tcB_{tj}|^2, \\
\\
(\cB'(\sigma_1) - \cB'(\sigma_2))(t) 
:= \sum_{i=1}^k \lambda_i |\braket{b'_i}{t}|^2 -
   \sum_{j=k+1}^{k+l} \mu_j |\braket{b'_j}{t}|^2 
 = \sum_{i=1}^k \lambda_i |\cB'_{ti}|^2 -
   \sum_{j=k+1}^{k+l} \mu_j |\cB'_{tj}|^2 
\end{array}
\]
respectively, where $1 \leq t \leq n$. 
We now have
\begin{eqnarray*}
&      &
\!\!\!\!\!
\!\!\!\!\!
\!\!
\totvar{\hcB(\sigma_1) - \hcB(\sigma_2)} 
    =    \totvar{\tcB(\sigma_1) - \tcB(\sigma_2)} 
    =    \sum_{t=1}^n 
         \left|
         \sum_{i=1}^k \lambda_i |\braket{\tb_i}{t}|^2 -
         \sum_{j=k+1}^{k+l} \mu_j |\braket{\tb_j}{t}|^2 
         \right| \\
& \geq & \sum_{t=1}^n 
         \left|
         \sum_{i=1}^k \lambda_i |\braket{b'_i}{t}|^2 -
         \sum_{j=k+1}^{k+l} \mu_j |\braket{b'_j}{t}|^2 
         \right| -
         \sum_{t=1}^n 
         \left|
         \sum_{i=1}^k \lambda_i 
         \left(|\braket{b'_i}{t}|^2 - |\braket{\tb_i}{t}|^2\right)
         \right| - \\
&      & \sum_{t=1}^n 
         \left|
         \sum_{j=k+1}^{k+l} \mu_j 
         \left(|\braket{b'_j}{t}|^2 - |\braket{\tb_j}{t}|^2\right)
         \right| \\
& \geq & \totvar{\cB'(\sigma_1) - \cB'(\sigma_2)} -
         \sum_{i=1}^k \lambda_i 
         \sum_{t=1}^n 
         \left|
         |\braket{b'_i}{t}|^2 - |\braket{\tb_i}{t}|^2
         \right| - 
         \sum_{j=k+1}^{k+l} \mu_j 
         \sum_{t=1}^n 
         \left|
         |\braket{b'_j}{t}|^2 - |\braket{\tb_j}{t}|^2
         \right| \\
& \geq & \totvar{\cB'(\sigma_1) - \cB'(\sigma_2)} -
         \sum_{i=1}^k \lambda_i 
         \trnorm{\ketbra{b'_i} - \ketbra{\tb_i}} -
         \sum_{j=k+1}^{k+l} \mu_j 
         \trnorm{\ketbra{b'_j} - \ketbra{\tb_j}} \\
& \geq & \totvar{\cB'(\sigma_1) - \cB'(\sigma_2)} -
         O\left(\frac{1}{K \sqrt{k+l}}\right) \cdot
         \sum_{i=1}^k \lambda_i -
         O\left(\frac{1}{K \sqrt{k+l}}\right) \cdot
         \sum_{j=k+1}^{k+l} \mu_j \\
&   =  & \totvar{\cB'(\sigma_1) - \cB'(\sigma_2)} -
         t \cdot O\left(\frac{1}{K \sqrt{k+l}}\right)
  \geq   \totvar{\cB'(\sigma_1) - \cB'(\sigma_2)} -
         O\left(\frac{f}{K}\right)  
\end{eqnarray*}
with probability at least 
$1 - \exp\left(-\Omega\left(\frac{\sqrt{n}}{K}\right)\right)$ 
over the choice of $\cB'$.
The third inequality follows from the fact that the trace
distance between two quantum states
is an upper bound on the total variation distance
between the probability distributions got by performing
a measurement on the two states.

We generate $\cB'$ by first generating an $n \times (k+l)$ matrix
$\cB$ whose
entries are independent complex-valued 
random variables whose real and imaginary parts are each
independently distributed according
to the Gaussian $\cG$, and then normalising each column of $\cB$ 
in order
to get $\cB'$. Let $b_1, \ldots, b_{k+l}$ denote 
the columns of $\cB$. 
Since 
$\exp(-\epsilon / 2) \cdot \sqrt{1 + \epsilon} \leq -\epsilon^2/3$ 
for $0 \leq \epsilon \leq 1/2$, using $\epsilon = f/10$ in
Fact~\ref{fact:chernoffchisquare} we see that with probability
at least $1 - (k+l) \exp(-\Omega(f^2 n))$ over the choice
of $\cB$, $\norm{b_i}^2 \leq 2 n \left(1 + \frac{f}{10}\right)$
for $1 \leq i \leq k$ and
$\norm{b_j}^2 \geq 2 n \left(1 - \frac{f}{10}\right)$ 
for $k+1 \leq j \leq k+l$.
Consider any fixed $t$, $1 \leq t \leq n$. By
Proposition~\ref{prop:berryesseen}, with 
probability at least $c^2$ over the choice of $\cB$, 
\[
\sum_{i=1}^k \lambda_i |\cB_{ti}|^2 
>    2 \sum_{i=1}^k \lambda_i + \sqrt{2 \sum_{i=1}^k \lambda_i^2}
\geq t + f
\quad
\mbox{{\rm and}}
\quad
\sum_{j=k+1}^{k+l} \mu_j |\cB_{ti}|^2 
< 2 \sum_{j=k+1}^{k+l} \mu_j 
= t.
\]
Call the above event $E_t$. 
If $E_t$ occurs we have
\begin{eqnarray*}
&      &
\!\!\!\!\!
\!\!\!\!\!
\!\!
\left|
\sum_{i=1}^k \lambda_i |\cB'_{ti}|^2 -
\sum_{j=k+1}^{k+l} \mu_j |\cB'_{tj}|^2 
\right|
  >   
\frac{t+f}{2 n (1 + \frac{f}{10})} - 
\frac{t}{2 n (1 - \frac{f}{10})} 
  =   
-\frac{t f}{10 n (1 - \frac{f^2}{100})} + 
\frac{f}{2 n (1 + \frac{f}{10})} \\
& > & 
\frac{f}{2 n} 
\left(\frac{1}{1 + \frac{\sqrt{2}}{10}} - \frac{2}{5}\right) 
  >
\frac{f}{6 n}.
\end{eqnarray*}
Since the events $E_t$ for different $t$ are independent, using a
standard
Chernoff bound, with probability at least $1 - \exp(-\Omega(n))$
over the choice of $\cB$, at least $\frac{c^2 n}{2}$ different
$t$ will satisfy the above inequality. This means that
with probability at least 
$1 - \exp(-\Omega(n)) - (k+l) \exp(-\Omega(f^2 n))$ 
over the choice of $\cB$, 
$\totvar{\cB'(\sigma_1) - \cB'(\sigma_2)} \geq \frac{f c^2}{12}$.
Thus, with probability at least 
$
1 - \exp(-\Omega(n)) - (k+l) \exp(-\Omega(f^2 n)) -
\exp\left(-\Omega\left(\frac{\sqrt{n}}{K}\right)\right) \geq
1 - \exp\left(-\Omega\left(\frac{\sqrt{n}}{K}\right)\right) -
\frac{\sqrt{n}}{K} \cdot \exp(-\Omega(f^2 n))
$ 
over the choice of 
a random orthonormal basis $\hcB$ of $\C^n$, 
$
\totvar{\hcB(\sigma_1) - \hcB(\sigma_2)} >
\frac{f c^2}{12} - O(f/K). 
$
Since $c$ is a universal constant, we can choosing $K$ to be a 
sufficiently large universal constant thus proving the first
part of the lemma.

We now proceed to the proof of the second part of the lemma.
Let $\lambda_1, \ldots, \lambda_k$ be the positive eigenvalues
and $-\mu_{k+1}, \ldots, -\mu_n$ the non-positive eigenvalues
of $\sigma_1 - \sigma_2$.
By symmetry, we can assume that we are working in the eigenbasis
of $\sigma_1 - \sigma_2$, i.\,e., the eigenbasis of
$\sigma_1 - \sigma_2$ is the computational basis.
Define the $n \times n$ matrix $\cB$ to be the matrix
whose column vectors are $b_1, \ldots, b_n$. 
Suppose $v$ is a unit vector in $\C^m$. Then,
\[
\qform{v}{\sum_{i=1}^m b_i b_i^\dag}{v} =
\sum_{i=1}^n |v^\dag b_i|^2 =
\norm{v^\dag \cB}^2 =
\norm{\cB^\dag v}^2.
\]
Hence we have 
\[
\ell = \norm{\sum_{i=1}^n b_i b_i^\dag} = 
\max_v \qform{v}{\sum_{i=1}^m b_i b_i^\dag}{v} =
\norm{\cB^\dag}^2 = 
\norm{\cB}^2,
\]
where the maximum is taken over all unit vectors $v \in \C^n$.
The second equality follows because $\sum_{i=1}^n b_i b_i^\dag$
is a positive matrix.
By Lemma~\ref{lem:normrandmatrix}, 
$\ell = \norm{\cB}^2 \leq O(n \log n)$
with probability at least $1 - \exp(-\Omega(n \log n))$ over
the choice of $\cB$. 

Now,
\begin{eqnarray*}
&      &
\!\!\!\!\!
\!\!\!\!\!
\!\!
\totvar{\cM(\sigma_1) - \cM(\sigma_2)} 
    =  
\frac{1}{\ell}
\sum_{t=1}^n 
\left|b_t^\dag \sigma_1 b_t - b_t^\dag \sigma_2 b_t\right| +
|\Tr (\sigma_1 \nu) - \Tr (\sigma_2 \nu)| \\
& \geq &
\frac{1}{\ell}
\sum_{t=1}^n 
\left|b_t^\dag \sigma_1 b_t - b_t^\dag \sigma_2 b_t\right| 
    =   
\frac{1}{\ell} \sum_{t=1}^n 
\left|
\sum_{i=1}^k \lambda_i |b_t^\dag \ket{i}|^2 -
\sum_{j=k+1}^n \mu_j |b_t^\dag \ket{j}|^2
\right| \\
& \geq &
\Omega\left(\frac{1}{n \log n}\right) \cdot \sum_{t=1}^n 
\left|
\sum_{i=1}^k \lambda_i |\cB_{it}|^2 -
\sum_{j=k+1}^n \mu_j |\cB_{jt}|^2
\right|.
\end{eqnarray*}
By Proposition~\ref{prop:berryesseen} and a standard Chernoff
bound, we see that with probability at least 
$1 - \exp(-\Omega(n))$
over the choice of $\cB$, for at least $\frac{c^2 n}{2}$
different $t$, 
\[
\left|
\sum_{i=1}^k \lambda_i |\cB_{it}|^2 -
\sum_{j=k+1}^n \mu_j |\cB_{jt}|^2 
\right|
  >    
2 \sum_{i=1}^k \lambda_i + \sqrt{2 \sum_{i=1}^k \lambda_i^2} -
2 \sum_{j=k+1}^n \mu_j 
 \geq
t+f - t 
   =
f.
\]
Thus, with probability at least
$
1 - \exp(-\Omega(n)) - \exp(-\Omega(n \log n)) \geq 
1 - \exp(-\Omega(n))
$
over the choice of $\cB$, 
\[
\totvar{\cM(\sigma_1) - \cM(\sigma_2)} 
> \Omega\left(\frac{1}{n \log n}\right) \cdot 
  \frac{c^2 f}{n} 
= \Omega\left(\frac{f}{\log n}\right), 
\]
since $c$ is a universal constant. 
Since the POVM $\cM$ can be refined to a POVM 
with $2n$ rank one elements, $\cM$ can be implemented as an 
orthonormal measurement in $\C^n \otimes \C^2$.
This completes the proof of
the second part of the lemma.
\end{proof}

We are now finally in a position to prove the main theorem of
the paper. 
\begin{theorem}
\label{thm:frob}
Let $\sigma_1$, $\sigma_2$ be two density matrices in $\C^n$.
Define $f := \frob{\sigma_1 - \sigma_2}$. Then:
\begin{enumerate}
\item
Let $K > 1$ be a sufficiently large quantity.
Consider an ancilla space $\C^m$ initialised to zero, where 
$m \geq \frac{4 n K^2}{f^2}$. 
Let $\hcB$ be a random orthonormal measurement basis in 
$\C^n \otimes \C^m$.
Let $\cM$ denote the POVM on $\C^n$ got by attaching ancilla
$\ket{0}$ to a state in $\C^n$ and applying the orthonormal
measurement $\hcB$ in $\C^n \otimes \C^m$. Then with probability
at least $1 - \exp(-\Omega(K n))$ over the choice of $\hcB$,
$\totvar{\cM(\sigma_1) - \cM(\sigma_2)} > \Omega(f)$; 

\item
Let $K \geq 1$ and define $m := K n$.
Take a set $\cB$ of $m$ independent random vectors 
$\cB := \{b_1, \ldots, b_m\}$ in 
$\C^n \otimes \C^K$, where each $b_i$ is got by choosing $m$ 
independent
complex numbers whose real and imaginary parts are independently
chosen according to the Gaussian $\cG$.  Define
$\ell := \norm{\sum_{i=1}^m b_i b_i^\dag}$ and
$
\nu := \onemat_{\C^n \otimes \C^K} - \frac{1}{\ell} 
       \sum_{i=1}^m b_i b_i^\dag.
$
Let $\cM$ denote
the POVM on $\C^n$ got by tensoring a zero ancilla over $\C^K$ to
states in $\C^n$ and then performing the POVM $\bcM$ in
$\C^n \otimes \C^K$ consisting of the elements 
$\frac{b_i b_i^\dag}{\ell}$ for $1 \leq i \leq m$,
and the element $\nu$. Note that $\cM$ can be implemented
as an orthonormal measurement in $\C^n \otimes \C^{2K}$.
Then with probability at least $1 - \exp(-\Omega(m))$ over the
choice of $\cB$, 
$
\totvar{\cM(\sigma_1) - \cM(\sigma_2)} >
\Omega\left(\frac{f}{\log m}\right).
$
\end{enumerate}
\end{theorem}
\begin{proof}
In order to prove the first part of the theorem,
let $K$ be at least as large as the universal
constant in the first part of Lemma~\ref{lem:frob}.
Thus, we start out with two density matrices 
$\bsigma_1 := \sigma_1 \otimes \ketbra{0}$, 
$\bsigma_2 := \sigma_2 \otimes \ketbra{0}$
in $\C^n \otimes \C^m$. Trivially, 
$
\rank{\bsigma_1} + \rank{\bsigma_2} =
\rank{\sigma_1} + \rank{\sigma_2} \leq 
2 n \leq 
\sqrt{n m}/K.
$
Also, $\frob{\bsigma_1 - \bsigma_2} = \frob{\sigma_1 - \sigma_2}$.
By the first part of Lemma~\ref{lem:frob}, with probability at least 
$
1 - \exp(-\Omega(\sqrt{n m})) -
\frac{\sqrt{n m}}{K} \cdot \exp(-\Omega(n m f^2))
\geq 1 - \exp(-\Omega(K n))
$ 
over the choice of 
a random orthonormal basis $\hcB$ of $\C^n \otimes \C^m$, 
$
\totvar{\cM(\sigma_1) - \cM(\sigma_2)} =
\totvar{\hcB(\bsigma_1) - \hcB(\bsigma_2)} >
\Omega(f). 
$
This completes the proof of the first part of the theorem.

A very similar strategy allows us to prove the second part of the
theorem using the second part of Lemma~\ref{lem:frob}.
\end{proof}
\paragraph{Remark:} The point to note in the second part of
the theorem is that the construction of the random POVM $\cM$
does not require a priori knowledge of 
$\frob{\sigma_1 - \sigma_2}$. This will be useful in the application
to the HSP, in the proof of Theorem~\ref{thm:randqft}

Finally, we present an example of a pair of density matrices
$\sigma_1$, $\sigma_2$ where with high probability
a random POVM cannot achieve
a total variation distance much larger than 
$\sqrt{\frob{\sigma_1 - \sigma_2}}$, unless the dimension of the
ancilla  used by the POVM is exponentially larger than 
$\rank{\sigma_1} + \rank{\sigma_2}$.
This is essentially because a sum of independent random variables
cannot deviate from its mean by much more than its standard
deviation. 
\begin{proposition}
\label{prop:highrank}
Let $\sigma_1$, $\sigma_2$ be completely mixed states supported on
two orthogonal $r$-dimensional subspaces of $\C^n$. 
Note that $\frob{\sigma_1 - \sigma_2} = \sqrt{2/r}$ and
$\trnorm{\sigma_1 - \sigma_2} = 2$.
Let $\cB$ be a random orthonormal basis in $\C^n$.
Then, with probability at least $1 - n \exp(-\sqrt{r})$ over the
choice of $\hcB$, 
$\totvar{\cB(\sigma_1) - \cB(\sigma_2)} \leq O(r^{-1/4})$.
\end{proposition}
\begin{proof}
Let $\cB = \{\ket{b_1}, \ldots, \ket{b_n}\}$. 
Let $W_1$, $W_2$ denote the supports of $\sigma_1$, $\sigma_2$
respectively. Then, $\sigma_i = \frac{1}{r} \Pi_{W_i}$.
Since each $\ket{b_t}$ is a random unit vector in $\C^n$,
putting $\epsilon = C r^{-1/4}$, $C$ a universal constant whose
value will become clear later, in the second part of
Lemma~\ref{lem:proj}, we get
$
\frac{1 - \epsilon}{n} \leq
\qform{b_t}{\sigma_i}{b_t} \leq
\frac{1 + \epsilon}{n}
$
for $i = 1, 2$ and all $1 \leq t \leq n$, with probability
at least $1 - n \exp(-\sqrt{r})$ over the choice of $\cB$.
Thus,
\[
\totvar{\cB(\sigma_1) - \cB(\sigma_2)} =
\sum_{t=1}^n
|\qform{b_t}{\sigma_1}{b_t} - \qform{b_t}{\sigma_2}{b_t}| \leq
\sum_{t=1}^n \frac{2 \epsilon}{n} \leq
2 \epsilon.
\]
This completes the proof of the proposition.
\end{proof}

Now, if we think of $\C^n$ as $\C^{2r} \otimes \C^{m}$, where
$m := \frac{n}{2r}$, we see that a random POVM in $\C^{2r}$
cannot distinguish
between $\sigma_1$, $\sigma_2$ by much more than
$\sqrt{\frob{\sigma_1 - \sigma_2}}$, unless $n$ is exponentially large
compared to $\rank{\sigma_1} + \rank{\sigma_2}$.

\section{Random measurement bases and the HSP}
In this section, we study the implications of Theorem~\ref{thm:frob}
for the hidden subgroup problem. 

Theorem~\ref{thm:frob} is in most cases
not immediately useful in
obtaining single register algorithms for the HSP. This is because
for two candidate hidden subgroups $H_1$, $H_2$,
$
\frob{\sigma_{H_1} - \sigma_{H_2}} \leq 
\frob{\sigma_{H_1}} + \frob{\sigma_{H_2}} =
\sqrt{\frac{|H_1|}{|G|}} + \sqrt{\frac{|H_2|}{|G|}}.
$
Thus, even though $\trnorm{\sigma_{H_1} - \sigma_{H_2}} \geq 1$ by
Proposition~\ref{prop:hsptrdist},
$\frob{\sigma_{H_1} - \sigma_{H_2}}$ can be exponentially small
if $|H_1|$, $|H_2|$ are exponentially small compared to
$|G|$. In most examples of interest this is indeed the case.
Fortunately, we can make good use of the fact that the coset states
for different subgroups of $G$ are simultaneously block diagonal
in the Fourier basis of $G$.  
Hence, we investigate the 
power of random Fourier sampling in distinguishing between
coset states. The advantage of this is that after doing the quantum
Fourier transform and measuring an irrep name and a row index,
we may be left with a reduced state on the space of column
indices with polynomially bounded rank. 
If this happens,
the average Frobenius distance between
the blocks of $\sigma_{H_1}$ and $\sigma_{H_2}$ will be polynomially
large even though $\frob{\sigma_{H_1} - \sigma_{H_2}}$ may be
exponentially small. 
In fact,
for several cases of the HSP studied in the literature,
the rank of the reduced state is in fact either $0$ or $1$ 
i.\,e., the hidden subgroup forms a Gel'fand pair with the
ambient group.

To make the above reasoning precise, we define a new distance metric
between two coset states $\sigma_{H_1}$, $\sigma_{H_2}$.
Below, we use the notation of Section~\ref{subsec:hspqft}.
\begin{definition}[$r(H_1, H_2)$]
\label{def:rh1h2}
Let $G$ be a group and $H_1, H_2 \leq G$. 
Define
\[
r(H_1, H_2) := w(H_1, H_2) + \frac{1}{{|G| \log |G|}} \cdot
\sum_{\rho \in \hG} d_\rho \, 
                    \frob{|H_1| \rho(H_1) - |H_2| \rho(H_2)}
\]
\end{definition}

The importance of $r(H_1, H_2)$ follows from the following
theorem.
\begin{theorem}
\label{thm:randqft}
Let $G$ be a group and $H_1, H_2 \leq G$. 
Let $\cM$ denote
the POVM corresponding to the following random Fourier sampling
procedure: apply $\QFT_G$ to the given coset state, measure the
name of an irrep $\rho \in \hG$ and a row index $i$, and then
apply a random POVM $\cM_\rho$ on the resulting reduced state
on the space of column indices, where $\cM_\rho$ is defined
as in the second part of 
Theorem~\ref{thm:frob} with 
$K_\rho := \ceil{\frac{C \log^2 |G|}{d_\rho}}$, where $C$ is a
sufficiently large universal constant. Then with probability at
least $1 - \exp(-\log^2 |G|)$ over the choice of $\cM$,
$
\totvar{\cM(\sigma_{H_1}) - \cM(\sigma_{H_2})} \geq 
\Omega(r(H_1, H_2)).
$
\end{theorem}
\begin{proof}
Let $\sigma_1$, $\sigma_2$ be two quantum states 
and $p_1, p_2 \geq 0$. Suppose $p_1 \geq p_2$. 
Then,
\[
\frob{p_1 \sigma_1 - p_2 \sigma_2} \leq
\frob{p_1 (\sigma_1 - \sigma_2)} + \frob{(p_1 - p_2) \sigma_2} \leq
p_1 \frob{\sigma_1 - \sigma_2} + |p_1 - p_2|.
\]
Now, 
\[
\totvar{p_1 \cM(\sigma_1) - p_2 \cM(\sigma_2)} =
\totvar{p_1 (\cM(\sigma_1) - \cM(\sigma_2)) + 
        (p_1 - p_2) \cM(\sigma_2)} \geq
\frac{p_1}{2} \totvar{\cM(\sigma_1) - \cM(\sigma_2)}.
\]
The inequality above follows by considering those outcomes of
$\cM$ that have at least as much probability for $\sigma_1$ as
for $\sigma_2$, and the fact that 
$(p_1 - p_2) \cM(\sigma_2)$ is a vector with non-negative entries.
Also, 
$\totvar{p_1 \cM(\sigma_1) - p_2 \cM(\sigma_2)} \geq |p_1 - p_2|$.
Now suppose 
$
\totvar{\cM(\sigma_1) - \cM(\sigma_2)} \geq
\frac{\frob{\sigma_1 - \sigma_2}}{L},
$
where $L \geq 1$.
Then,
\begin{eqnarray*}
&      &
\!\!\!\!\!
\!\!\!\!\!
\!\!
\totvar{p_1 \cM(\sigma_1) - p_2 \cM(\sigma_2)} 
  \geq  
\frac{|p_1 - p_2|}{2} + 
\frac{p_1}{4} \totvar{\cM(\sigma_1) - \cM(\sigma_2)}  \\
& \geq &
\frac{|p_1 - p_2|}{4 L} + 
\frac{p_1}{4 L} \frob{\sigma_1 - \sigma_2} 
  \geq
\frac{\frob{p_1 \sigma_1 - p_2 \sigma_2}}{4 L}.
\end{eqnarray*}

Now suppose we apply $\QFT_G$ and measure an irrep name
$\rho$ and a row index $i$. We apply the above reasoning 
to the random POVM $M_\rho$ with $L = \log |G|$. Using the
second part of Theorem~\ref{thm:frob}, we get that
with probability at least $1 - \exp(-\log^2 |G|)$ over
the choice of $M_\rho$, 
$
\totvar{\cM_\rho(\rho(H_1)) - \cM_\rho(\rho(H_2))} \geq
\Omega
\left(
\frac{\frob{|H_1| \rho(H_1) - |H_2| \rho(H_2)}}{\log |G|} 
\right).
$
Hence for the random Fourier sampling POVM $\cM$, 
with probability at least $1 - \exp(-\log^2 |G|)$ over
the choice of $\cM$,
\[
\totvar{\cM(\sigma_{H_1}) - \cM(\sigma_{H_2})} \geq
\Omega
\left(
\frac{1}{{|G| \log |G|}} \cdot
\sum_{\rho \in \hG} d_\rho \, \frob{|H_1| \rho(H_1)-|H_2| \rho(H_2)}
\right).
\]
The theorem now follows because
random Fourier sampling always does at least as well
as weak Fourier sampling.
\end{proof}

The following corollary is now easy to prove.
\begin{corollary}
\label{cor:randqft}
Let $G$ be a group. Suppose 
for every irrep $\rho \in \hG$ and subgroup $H \leq G$,
$\rank{\rho(H)} \leq (\log |G|)^{O(1)}$. Then the
random Fourier method of Theorem~\ref{thm:randqft}
gives rise to a single register algorithm
identifying 
with probability at least $3/4$ the hidden subgroup $H$
from $(\log |G|)^{O(1)}$ copies of $\sigma_H$.
\end{corollary}
\begin{proof}
Consider two distinct subgroups $H_1, H_2 \leq G$.
Since coset states are block diagonal in the Fourier basis of $G$,
using 
Theorem~\ref{thm:randqft}, Proposition~\ref{prop:hsptrdist} and 
Fact~\ref{fact:trnormfrobnorm} we get
\begin{eqnarray*}
&      &
\!\!\!\!\!
\!\!\!\!\!
\!\!
1 
  \leq   \trnorm{\sigma_{H_1} - \sigma_{H_2}}
    =    \sum_{\rho \in \hG} \frac{d_\rho}{|G|}
         \trnorm{|H_1| \rho(H_1) - |H_2| \rho(H_2)} \\
& \leq & \sum_{\rho \in \hG} \frac{d_\rho}{|G|}
         \frob{|H_1| \rho(H_1) - |H_2| \rho(H_2)} \cdot
         (\rank{\rho(H_1)} + \rank{\rho(H_2)}) \\
& \leq & (\log |G|)^{O(1)} \cdot
         \left(
         \sum_{\rho \in \hG} \frac{d_\rho}{|G|}
         \frob{|H_1| \rho(H_1) - |H_2| \rho(H_2)} 
         \right) \\
& \leq & (\log |G|)^{O(1)} \cdot r(H_1, H_2).
\end{eqnarray*}
Let $\cM$ denote the random Fourier sampling POVM of 
Theorem~\ref{thm:randqft}. Then with probability at least
$1 - \exp(-\log^2 |G|)$ over the choice of $\cM$,
$
\totvar{\cM(\sigma_{H_1}) - \cM(\sigma_{H_2})} \geq 
\Omega(r(H_1, H_2)) \geq
(\log |G|)^{-O(1)}.
$
Since a group $G$ can have at most $2^{\log^2 |G|}$ subgroups,
by the union bound on probabilities, with probability at
least $1 - \exp(-\Omega(\log^2 |G|))$ over the 
choice of $\cM$,
$
\totvar{\cM(\sigma_{H_1}) - \cM(\sigma_{H_2})} \geq 
(\log |G|)^{-O(1)}
$
for all subgroups $H_1, H_2 \leq G$. The corollary now follows
from Fact~\ref{fact:minfinding}.
\end{proof}

Finally, we remark that in many important examples of the HSP where
most of the probability lies on high dimensional irreps and the
blocks corresponding to these irreps have low rank,
one can save a factor of $\log |G|$ in the denominator of
the definition of $r(H_1, H_2)$ and prove Theorem~\ref{thm:randqft}
with this improved definition of $r(H_1, H_2)$. 
This improvement follows by using
the first part of Lemma~\ref{lem:frob} instead of the second part
of Theorem~\ref{thm:frob} in the proof of Theorem~\ref{thm:randqft}.
Such a saving can be done, for example, for suitable
subgroups of the affine group, Heisenberg group and
groups $\Z_p^r \rtimes \Z_p$, $p$ prime, $r \geq 2$.

\section{The general state identification problem}
In this section, we study the implications of Theorem~\ref{thm:frob}
to the state identification problem for a general ensemble of
quantum states. To the best of our knowledge, this problem does not
seem to have been studied before. The following theorem
with $r = n$ gives an upper bound on
the number of copies required to identify a given state 
information-theoretically with high probability for any
ensemble.
\begin{theorem}
\label{thm:genidentify}
Let $\cE = \{\sigma_1, \ldots, \sigma_k\}$ be an a priori known
ensemble of quantum states in $\C^n$. Suppose the minimum trace
distance between a pair of states from $\cE$ is at least $t$.
Let $r$ denote the maximum rank of a state in $\cE$.
Then, there is a POVM $\cM$ in $\C^n$ such that
$\cM^{\otimes \ell}$ acting on $\sigma_i^{\otimes \ell}$ gives enough
classical information to identify $i$ with probability
at least $3/4$, 
where $\ell = O\left(\frac{r \log k}{t^2}\right)$. 
\end{theorem}
\begin{proof}
Define $f := \frac{t}{\sqrt{r}}$.
Let $\cM$ be the random POVM guaranteed by the second part of 
Theorem~\ref{thm:frob} with 
$m := \frac{16 n K^2 \log^2 m}{f^2}$. 
Fix any pair of states $\sigma_i$, $\sigma_j$, $i \neq j$ from $\cE$.
Then with probability at least 
$1 - \exp(-\Omega(8 n \log m)) \geq 1 - \frac{1}{m^2}$ over 
the choice of $\cM$,
\[
\totvar{\cM(\sigma_i) - \cM(\sigma_j)} > 
\Omega(\frob{\sigma_i - \sigma_j}) \geq
\Omega\left(\frac{\trnorm{\sigma_i - \sigma_j}}
                 {\sqrt{\rank{\sigma_i - \sigma_j}}}
      \right) \geq
\Omega\left(\frac{t}{\sqrt{r}}\right).
\]
By the union bound on probabilities, there is a POVM $\cM$ on
$\C^n$ such that the above inequality holds for every pair
of states from $\cE$. By Fact~\ref{fact:minfinding},
applying $\cM^{\otimes \ell}$ on $\sigma_i^{\otimes l}$,
where $\ell = O\left(\frac{r \log k}{t^2}\right)$ gives 
enough classical information to identify $i$ with probability
at least $3/4$.
\end{proof}


\subsection*{Acknowledgements}
The author wishes to thank Andris Ambainis,
Martin R\"{o}tteler, Debbie Leung, Joseph
Emerson and Christoph Dankert for
useful discussions, and Jaikumar Radhakrishnan for feedback on
an earlier version of the paper.

\bibliographystyle{alpha}
\bibliography{frobenius}

\end{document}